# Ethics, Information, and Our "It-from-Bit" Universe

**Terrell Ward Bynum**
**Southern Connecticut State University**

The essence of the Computer Revolution is found in the nature of a computer itself. What is revolutionary about computers is *logical malleability*. *James H. Moor 1985*

It from bit . . . every particle, every field of force, even the spacetime continuum itself . . . derives its function, its meaning, its very existence from [bits]. *John Archibald Wheeler 199*

**Abstract:** Using information technology, humans have brought about the "Information Revolution," which is changing the world faster and more profoundly than ever before, and generating an enormous number of ethical "policy vacuums". How is this possible? An answer is suggested by ideas from James Moor regarding "logical malleability," in his classic paper "What is Computer Ethics?" (1985) The present essay combines Moor's ideas with the hypothesis that all physical entities — including spacetime and the universe as a whole — are dynamic data structures. To show the usefulness of taking such an approach, in both physics and in computer ethics, a suggested "it-from-bit" model of the universe is briefly sketched, and relevant predictions are offered about the future of computer and information ethics.

## *1. Introduction*

One of the most fruitful and insightful claims about the nature of the universe was made by philosopher/scientist Norbert Wiener in an address to the New York Academy of Sciences in autumn of 1946. Apparently, Wiener was the first to recognize the physical nature of "environmental information," as it is often called today. It suddenly became clear to him — while advising Claude Shannon on how to determine the amount of information being carried in a telephone wire (Conway and Siegelman, 2005, 164, 186) — that *environmental information is physical and is fundamentally related to the second law of thermodynamics*. Wiener realized that such information plays a major role in every physical entity and process. Delighted by this insight, he walked the halls of his MIT building, in one of his "Wiener walks" (Conway and Siegelman, 2005, Ch 10), telling everyone he met that entropy is a measure of information (Rheingold, 1985, 112).

Wiener's new understanding of entropy, gave physicists and philosophers alike a new account of the nature of physical objects and processes — and, indeed, *a new way to understand the ultimate nature of the universe* (see Section 7 below). He saw that all physical entities are "informational objects" or "informational processes." Even living beings are entities that store information in their genes and use it to create amino acids and proteins, and even to create new living beings — their own offspring. In addition, the nervous systems of animals receive, store and process physical information, thereby generating animal activities, perceptions, and emotions. Human beings also, Wiener noted, can be described as *informational entities* — persisting patterns of information:

> We are but whirlpools in a river of ever-flowing water. We are not stuff that abides, but patterns that perpetuate themselves. *Wiener 1954, 96*
>
> . . .
>
> The individuality of the body is that of a flame . . . of a form rather than of a bit of substance. *Wiener 1954, 102*

Significant developments in physics after Wiener have deepened and extended this "Wienerian" understanding of the cosmos. For example, influential physicist John Archibald Wheeler famously said:

> It from bit . . . every particle, every field of force, even the spacetime continuum itself . . . derives its function, its meaning, its very existence from [bits] . . . . Tomorrow we will have learned to understand and express *all* of physics in the language of information. *Wheeler 1990, 310, 313, (*emphasis in the original)

Wheeler's famous "it-from-bit" declaration has led to significant growth in physics during the past three decades (examples discussed below).

Inspired by Wiener's insight, Wheeler's "it-from-bit" prediction, and Moor's "logical malleability" approach to computer ethics (Moor 1985), the present essay explores the idea that the ability of informational technology to represent, analyze and manipulate environmental information — even at the "deepest level" of physical existence — can cast significant light upon the rapid and far reaching changes made possible by information technology (IT), as well as the enormous number of ethical challenges (Moor calls them "policy vacuums") that arise from new IT-enabled capabilities.

To provide groundwork for the main discussion below, Sections 2, 3, and 4 introduce some preliminary ideas — Section 2 briefly summarizes key ideas from Moor's account of computer ethics; Section 3 discusses the concept of "physical information"; and Section 4 describes Plato's account of the structure of physical objects beneath the moon. After that, Sections 5 through 7, discuss several topics from contemporary physics — quantum information, nanotechnology, spacetime as a quantum data structure, and the universe as a single dynamic data structure. These ideas suggest an "it-from-bit" model of the universe, which is sketched in Section 7. Finally, Section 8 offers some predictions about the future of computer and information ethics in light of the current rapid growth in information physics.

## *2. Moor on Computer Ethics*

In Moor's account of Computer Ethics, he identifies what he calls "logical malleability" as the key to the revolutionary capacity of information technology to change the world:

> The essence of the Computer Revolution is found in the nature of a computer itself. What is revolutionary about computers is *logical malleability*. Computers are logically malleable in that they can be shaped and molded to do any activity that can be characterized in terms of inputs, outputs and connecting logical operations. . . . Because logic applies everywhere, the potential applications of computer technology appear limitless. . . . Indeed, the limits of computers are largely the limits of our own creativity. . . . The logic of computers can be massaged and shaped in endless ways through changes in hardware and software . . . . Syntactically, the logic of computers is malleable in terms of the number and variety of possible states and operations. Semantically, the logic of computers is malleable in that the states of the computer can be taken to represent anything. *Moor 1985, 269–70 (*italics in the original)

This power and flexibility of computers and related technology, according to Moor, make them nearly universal tools which can enable human beings to do many things that were never done before. But if they were never done before, there may be no laws, and no standards of good practice, and no other policies to determine whether it would be *morally right* to do them. Just because we *can* do them, doesn't mean that it would be *ethical* to do them. If the universe is, essentially, an "it-from-bit" data structure, even at the deepest level of physical existence, then the logical malleability of information technology will continue to enable people to do a vast number of things that generate new policy vacuums. But *is* the universe actually a huge "it-from-bit" data structure? The present essay is focused upon this question and possible consequences for the future of both physics and ethics. (See Sections 7 and 8)

## *3. What Is physical Information?*

Physical information consists of *physical data*, which is *syntactic*, not semantic. But what is a *datum*? As Luciano Floridi explains, a datum is essentially a difference:

> a datum is ultimately reducible to *a lack of* uniformity. More formally, according to the *diaphoretic interpretation* (*diaphora* is the Greek word for 'difference'), the general definition of a datum is:

> **Dd** datum =def. x being distinct from y
>
> where the x and the y are two uninterpreted variables and the domain is left open to further interpretation   *Floridi 2011, 85*

A physical datum is a difference "embodied in," "carried by" — some would say "encoded in" or "registered by" — the structure of a physical entity. All the differences embodied within a given physical entity, right down to the subatomic differences at the deepest level of existence (see Sections 5 and 6 below), constitute the physical structure of that entity. *So physical entities are data structures, most of them enormously complex; but the data which they encode are not matter-energy — the data are physical relations, not material objects.*

At the Big Bang, everything in the universe consisted of primordial, extremely hot energy. As the universe cooled, four fundamental forces of nature (or maybe *five* —see Section 7 below) assembled that energy into a wide variety of structures, including spacetime, elementary particles, and eventually every physical thing in the universe. So the fundamental forces of nature created information by assembling all the different structures that now comprise our universe. ("It from bit!")

Some differences embodied within physical entities are perceivable at the "macro-level," while others are non-perceivable differences existing at various "micro-levels." Eliminating physical differences within an entity — "erasing data" — erodes that entity's data structure. And if there is enough erosion, or the right sort of erosion, an entity can be significantly changed, damaged, or even destroyed. The loss of information embodied within a physical entity constitutes a loss of order and structure, which is what entropy measures.

Consider an everyday "macro-level" example: What makes a house a house? Spacetime relationships among various parts of the house comprise that portion of the house's data structure which *makes the house a house.* So, initially, a builder may begin with a pile of lumber, a pile of bricks, a pile of pipes, and so on. However, such piles do not constitute a house because they do not have the *form* of a house — that is, *they do not embody/carry/encode the appropriate spacetime relations.* However, when the builder uses boards, bricks, pipes, and so on, to build a house — thereby *changing the spacetime relations* among the various building supplies — those "supplies" become a house. So it is the *form* of the house, the *pattern of spacetime relations*, the *spacetime data structure*, that makes the house a house. The same can be said about every piece of building supplies from which the house was built — each board, each brick, each pipe, and so on, has a data structure which makes it a board or a brick or a pipe. W*ith appropriate quantum considerations* (see Section 5 below)*, one can even say the same thing for every atom and subatomic particle comprising a house* — they, too, are physical data structures*.* A house, therefore, is an extremely complex, continually changing (at least at the quantum level) data structure composed of a staggering number of smaller data structures (boards, bricks, pipes, molecules, atoms, particles, etc.). Recent developments in physics have led a number of scientists to believe that *even spacetime itself has a data structure* (See Section 6). If that is correct, *the entire universe, too, is a single, continually changing, data structure.* (See Section 7).

## *4. Plato's "Beautiful" Mathematical Data Structures*

A related metaphysical (and, in some circumstances, scientific) question is: What is *the source* of the order and structure found in our universe? An early answer was offered by Plato, who sought to explain the regularities and mathematical patterns — even the beauty! — of nature. (See, especially, Vlastos, 1975 and also Bynum, 2016) Because he lived more than two millennia ago, long before today's Information Revolution, Plato did not couch his metaphysical ideas and explanations in informational terms. Nevertheless, it may be of interest here to discuss, from an informational point of view, Plato's account of the nature of physical entities on or near the earth — "below the moon," as he would say. Plato accepted, from the philosopher Empedocles, the view that there are four basic elements (Empedocles called them "roots".) which comprise all physical entities: earth, air, fire and water. In addition, Plato accepted the hypothesis of the atomists, Leucippus and Democritus, that the matter comprising physical entities consists of very tiny, invisible units — "atoms."

In the second division of his *Timaeus* (47E–69B), Plato presents a description of the deep structure of matter, which, he says, was impressed upon "inchoate" matter by a totally rational god, the Demiurge (the "Craftsman" or "Artist"). As Gregory Vlastos explains, in his book, *Plato's Universe,*

> The matter which confronts the Demiurge in its primordial state is inchoate. The four primary kinds of matter, earth, water, air and fire, are present here in a blurred, indefinite form; their motion is disorderly. The Demiurge changes all this. He transforms matter from chaos to cosmos by impressing on it regular stereometric form. When he has done his job, all of the fire, air, and water in existence will consist of tetrahedra, octahedra, and icosahedra, respectively, that is to say of solids whose faces are invariably equilateral triangles. And earth will be found to consist of minute cubes [that is, hexahedra, whose faces are all squares]. *Vlastos 1975 and 2005, 69–70*

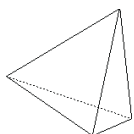 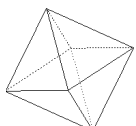 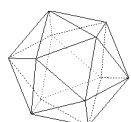 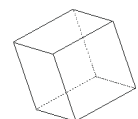

Tetrahedron       Octahedron       Icosahedron       Hexahedron

The Demiurge did not create inchoate matter. It was present before he "impressed" order and structure onto it, thereby "encoding information in the matter" (to use today's informational language) thus creating mathematically "beautiful" data structures. Accordingly, for Plato the fundamental structures beneath the moon are tetrahedra (fire), octahedra (air), icosahedra (water), and hexahedra (earth). Because all of these can be constructed from triangles, an informational description today would say that the triangles constitute an "even deeper microstructure" within earth, air, fire, and water. And because triangles, in turn, can be constructed from straight lines and angles, the lines and angles would comprise "the deepest microstructure" encoding physical information in entities beneath the moon.

The geometric "atoms" comprising earth, air, fire and water, according to Plato, are mere *imperfect copies* of the perfect, unchanging geometric forms which a rational mind can access in the realm of Platonic forms. Those perfect forms are *not part* of the imperfect physical earth, air, fire and water, according to Plato; but they can help a rational mind better understand the beautiful, orderly atoms comprising physical objects beneath the moon.

Plato's metaphysical explanation of the source of order and structure in the physical world has not worked out well from the point of view of contemporary science. It nevertheless reveals Plato's insight into the need to explain regular patterns and structures found in nature, as well as the remarkable ability of mathematics to reveal significant features of those structures.

## *5. Quantum Information and Nanotechnology*

"Classical" (that is, non-quantum) bits of information, like those stored and processed within traditional digital computers, are *binary* — they exist in either one of two states, *but not in both at the same time*. A vast number of physical objects and processes include data structures that can be represented within binary computers, making it possible to create, analyze, manipulate and control such objects and processes *in endless new ways.* Our ability to do this generates a continuous supply of new "policy vacuums" and ethical challenges (as Moor would say). So, apparently, the world will need computer ethicists well into the future!

In addition to containing classical bits, the data structures of most physical entities also contain *quantum bits — qubits* — which are significantly different from classical bits. Qubits are encoded by two-state quantum-mechanical systems, such as the polarity of a photon (vertical and horizontal polarization) or the "spin" of an electron ("spin-up" and "spin-down"). When an electron, for example, is not interacting with classical physical entities, its "spin," on a given axis, exists in *quantum superposition*; that is, its "spin" is *both "up" and "down" at the same time, and even simultaneously in a vast number of superpositions between "up" and "down."* However, when an electron in superposition is measured by an experimenter, or when such an electron interacts with classical physical objects, its quantum "spin," on a given axis, instantly "decoheres" — randomly — as a classical bit which is either "spin up" or "spin down," but not both. Similarly, the same decoherence phenomenon, from superpositions to classical bits, occurs in other quantum entities. Essentially, qubits embody information encoded as "clouds" or "waves" of physical possibilities that randomly "decohere" and become classical bits when a quantum entity interacts with another physical object.

During the past few decades, on the cutting edges of information theory, efforts to understand and make use of quantum information have been increasingly successful. The resulting *Quantum Information Theory* includes rules, theorems and procedures that, despite some similarities, are significantly different from Classical Information Theory (see, for example, Vedral 2006). Today, much progress has been made with regard to understanding and manipulating quantum phenomena like superposition, entanglement, and teleportation.  In addition, early versions of quantum computers have been created, tested, and marketed; and, indeed, MIT quantum computer engineer Seth Lloyd has argued that the universe itself can be considered to be an enormous quantum computer:

> The conventional view is that the universe is nothing but elementary particles. That is true, but it is equally true that the universe is nothing but

> bits — or rather nothing but qubits. Mindful that if it walks like a duck and it quacks like a duck then it's a duck . . . since the universe registers and processes information like a quantum computer, and is observationally indistinguishable from a quantum computer, then it *is* a quantum computer (italics in the original). *Lloyd, 2006, p. 154*

The Information Revolution has now advanced to the point where data structures of physical objects — including everyday perceivable differences all the way down to imperceivable "nano level" differences — can be identified, explained, manipulated and exploited. At the nano level, chemical reactions take place, molecules are formed, and many fundamental physical structures first appear. Thus, as Mihail Roco has noted:

> nanotechnology allows us to work at the very foundation of matter, the first level of organization for both living and man-made systems. At this scale all fundamental structures, properties, and functions of materials and devices are established. This is the lowest scale for transforming capabilities (manufacturing) for practical uses. *Roco, 2007, xi*

To work at the nano level is to manipulate and exploit *extremely tiny data structures*. A *nanometer*, for example, is one billionth of a meter. Entities so small are difficult to imagine, so Adam Keiper has offered the following helpful — and very striking — comparison:

> If a nanometer were somehow magnified to appear as long as the nose on your face, then a red blood cell would appear the size of the Empire State Building, a human hair would be about two or three miles wide, one of your fingers would span the continental United States, and a normal person would be about as tall as six or seven planet Earths piled atop one another. *Keiper, 2003, 18* (quoted in Allhoff, et al., 5)

So physicists now can manipulate and exploit data structures all the way down to the level where fundamental properties and functions of matter come into existence. The social and ethical implications of such power over nature — for biomedicine, agriculture, environmental protection, energy production, warfare, and so on (to name only a few examples) — have just begun to be identified and understood. As a result, the need for significant developments in Nano Information Ethics is growing rapidly (see, especially, the anthology *Nanoethics*, edited by Alloff, Lin, Moor and Weckert, 2007).

## *6. Does Spacetime Have a Data Structure?*

To identify and, eventually, come to understand *all* of the data structures of the universe, one must look even "deeper" than the nano level. In particular, the question of whether spacetime itself has a data structure needs to be answered. *If even* *spacetime has a data structure, then apparently the entire universe is a single, enormous, dynamic informational entity.* Given the powerful tools of the Information Revolution, such as classical information theory, quantum information theory, artificial intelligence, and quantum computing, humanity in the future will surely face a staggering number of policy vacuums and resulting ethical challenges.

A number of physicists working to determine whether space has a quantum data structure have recently made notable progress. One of the most promising approaches appears to be that of Loop Quantum Gravity which, if successful, would actually combine General Relativity and Quantum Mechanics into a single theory. (See especially Rovelli, 2017, Part Three, where promising, though not completely confirming, evidence is discussed regarding measurements of the cosmic background radiation and explanations of black hole evaporation.)

Efforts to develop "loop theory" — even before that name was used — began in 1966 when John Wheeler and Bryce DeWitt developed what came to be called the Wheeler-DeWitt equation based upon DeWitt's suggested "wave function of space." Since then a number of scientists have worked with the Wheeler-DeWitt equation and revisions of it. In the late 1980s, Abhay Ashtekar simplified the equation, and soon thereafter Lee Smolin and Ted Jacobson found the first solutions, which were all "loops" — that is, lines that close upon themselves. This is how Loop Quantum Gravity Theory got its name. (Rovelli, 2017, Chapter 5)

By the mid 1990s a "spectrum of the volume of space" had been developed by loop theorists, indicating that space does indeed have a quantum data structure. (Rovelli, 2017, 164-5) Thus, if Loop Quantum Gravity provides a correct analysis of the structure of space (not yet confirmed, though recently acquired evidence looks promising) *there is a smallest volume of space, and space exists as quantized "packets."* Such packets are not *in* space, they are the components *of* space itself. According to loop theory, the tiniest "grains" of space combine to form curved triangles, which can be represented by nodes and lines on a graph. As Rovelli explains:

> the key to understanding the physics of [loop theory] solutions lies in the points where these lines intersect. These points are called "nodes," and the lines between nodes are called "links." A set of intersecting lines is called a "graph," that is to say, a combination of nodes connected by links . . . . without nodes, physical space has no volume. In other words,

> it is in the nodes of the graph, not in the lines, that the volume of space "resides." The lines "link together" individual volumes sitting at the nodes. *Rovelli, 2017, 162-3*

So according to Loop Quantum Gravity, *space does have a data structure*, and that structure exists at the Planck level, where the smallest possible entities in the universe reside. (A Planck length is less than a billionth of a billionth of the diameter of a proton.) The data structure of space, according to loop theory, is created from curved triangles which combine to form the curvature of space — essentially, the gravity of Einstein's General Relativity Theory. So it is clear that recent achievements in Loop Quantum Gravity have been impressive, and early efforts to confirm the overall theory are promising. (See especially Rovelli, 2017, Ch 9)

## *7. New Models of the Universe?*

If one assumes that Loop Quantum Gravity eventually will be confirmed, a number of significant new research opportunities in physics will be generated; and, in addition, computer and information ethics will continue to have significant importance, because an enormous number of policy vacuums will result. In physics, for example, promising new models of the universe might be proposed with features of interest like these:

a. General Relativity and Quantum Mechanics would be combined into one consistent theory, thanks to Loop Quantum Gravity.

b. A new particle, which generates spacetime — perhaps aptly called a "spaton" — could be proposed to explain cosmic inflation, dark energy, and the cosmological constant. (See explanations below.)

c. A fifth fundamental force in the universe — perhaps appropriately named the "super strong force" — could be required to explain how spatons could stick together in spite of the enormous heat of the Big Bang.

d. A newly identified "cosmological arrow of time" — with two "clocks" to measure it — may put universe-wide time back into physics.

e. Primordial sources of several "constants of nature" might be identified for the first time, including the speed of light, the gravitational constant, the Planck constant, and the cosmological constant.

Consider, for example, what one might call *"The Spatons Model of the Universe"*:

*The Existence of "Spatons"* — Loop Quantum Gravity has provided promising evidence that spacetime has a data structure of extremely small quantum "packets" that combine to create curved space. The tiniest "packets" are *nodes* where units of spacetime reside, while *links* connecting the nodes are junctions where spacetime units stick together. (Rovelli 2017, Chapter 5) It seems reasonable to assume that, if all other entities in the universe consist of quantum particles, then spacetime also must consist of quantum particles. (Occam's Razor recommends that new kinds of entities should *not* be assumed unless they are absolutely necessary to account for the evidence. So taking spatons to be very tiny quantum particles, instead of some new kind of entity, makes a lot of sense.) The presumed spaton particles must to be able to combine during the extreme heat of the Big Bang. None of the particles of the Standard Model of Quantum Mechanics, however, can combine at such high temperatures, so the spaton would be a newly identified quantum particle.

*The Existence of a "Super Strong" Force* — According to current models of the Big Bang, spacetime came into existence about a trillionth of a second after the primordial energy plasma began to cool. Temperatures were very, very high, so *the force holding the tiny units of spacetime together must be remarkably strong;* otherwise, spacetime could never have come into existence. It follows that a fifth fundamental force (a "super strong" force) is needed to explain the data structure of spacetime. A related question is whether there exists another particle — one that might be called, say, a "*spagon*" — which carries the super strong force and "glues" spatons together, thereby creating spacetime and preventing it from disintegrating.

*Explaining Cosmic Inflation* — In current models of the Big Bang, almost immediately after the primordial plasma began to cool, the universe underwent an enormous expansion — cosmic inflation. A reasonable explanation of how this could have happened would be that, as cooling began, the primordial energy plasma entered a "phase change," much like steam does when it condenses into liquid water. So a trillionth of a second or so after the Big Bang started, vast quantities of spatons (and perhaps spagons) must have condensed out, instantly combining to create cosmic inflation.

*Explaining Dark Energy and the Cosmological Constant* — The current account of the "vacuum" of space is that, in reality, it is not actually empty. Instead, it contains "quantum foam," which is constantly swarming with "virtual particles" that come into existence for a tiny fraction of a second and then vanish as they annihilate each other.

It is also possible, however, that not all "vanishing" occurs because of annihilation? *If part of quantum foam consists of virtual spatons (and maybe virtual spagons), they could instantly combine, and thereby "vanish", becoming new spacetime.* The result would be continuous expansion of the universe, occurring faster and faster as more and more spacetime comes into existence. "Dark Energy," then, would turn out be simply new spacetime constantly being created everywhere within quantum foam; and the Cosmological Constant would remain constant because of the rate at which this process occurs.

*Simultaneity and the Cosmic "Arrow of Time"* — If one assumes that the above accounts of Dark Energy and the Cosmological Constant are essentially true, then cosmic time exists and has been flowing ever "forward" since inflation began. If so, there would be two "cosmic clocks" from which, in principle, one could tell "what time it is in the universe": the growing radius of the universe (assuming that it is spherical) and the growing volume (whatever the shape of the universe happens to be). Cosmic time would be "ticking" everywhere within quantum foam. The relativity of "local time," as described by Einstein's theory, would continue to be correct; but, for any two events that took place in different parts of the universe, the "cosmic clocks" would, in principle, be able to tell you whether or not they are simultaneous in cosmic time. If the cosmic radius or cosmic volume are the same size during the two events, then they are cosmically simultaneous. Simultaneity, therefore, would no longer be banished from the universe. (And Newton would turn out to be right that a universe-wide time, in every tiny location in the cosmos, would continually flow from the past into the future.)

*Explaining the Earliest Quasars and Giant Black Holes* — Quasars are remarkably bright entities caused by gas entering enormous black holes. According to current explanations, such black holes would require about a billion years to form. Nevertheless, recent observations indicate that giant black holes and the resulting quasars existed less than 700 million years after the Big Bang. How could they have developed so quickly? A number of physicists are currently working to solve this mystery. If Loop Quantum Gravity is correct, a reasonable explanation is available: since *spacetime is made of energy*, and since black holes consume everything that is made of energy which passes through their event horizons, it follows that black holes must consume spacetime. So "small" black holes, formed by "dying" first-generation stars, could grow not only by consuming gas, dust and other stars, they also could grow by consuming any spacetime that happens to go through their event horizons and thereby grow into giant black holes earlier than currently expected.

*A New Account of the End of the Universe* — If black holes consume spacetime, the following is at least possible: In the current era of cosmic time, existing black holes are not sufficient to consume spacetime faster than quantum foam creates new spacetime. The universe, therefore, is currently expanding. Many billions of years from now, however, as trillions upon trillions of stars have grown old and have become black holes themselves, they may be able to consume spacetime faster than quantum foam can create it. Thus, cosmic expansion would end, and the universe would become continually smaller and smaller. Eventually, all the black holes would combine to create a near-singularity which is so hot that it explodes and becomes a new Big Bang. A new universe would come into existence. Here we have yet another of the many suggested ways that the universe might "end" — and an optimistic view that new universes will come into existence using energy from the current one.

*Further Considerations* — If the spatons model of the universe, as sketched here is correct, then the spatons-generated data structure of spacetime might have determined the smallest possible length, area and volume that could exist in the universe (the Planck length, the Planck area and the Planck volume), as well as the curvature of spacetime (the gravity of General Relativity). In addition, the resulting data structure of spacetime, together with its electromagnetic properties, may have set the maximum speed at which light may travel through "empty" space. An additional possibility is that the data structure of spacetime could, perhaps, determine the allowable superpositions of quantum entities. For example, an electron traveling through spacetime as a wave of physical possibilities might only be able to take superposition paths made possible by the data structure of the particular local area of spacetime through which it is currently traveling.

The Spatons Model suggested here is just an example of an enormous number of suggestions and ideas that likely will emerge if Loop Quantum Gravity is confirmed. Loop Theory would be an exciting achievement in physics, and it would have significant implications for the importance and development of Computer and Information Ethics.

## *8. Conclusion — Computer Ethics and Our "It-from-Bit" Universe*

Given all that has been said above, and also assuming that Loop Quantum Gravity indeed will get confirmed, it is reasonable to conclude that *the universe is a single constantly changing informational structure.* Because of the logical malleability of computer and information technology — its ability to represent, model and manipulate just about anything — humanity will continue to be faced with a steady stream of new ethical "policy vacuums". This is already evident from recent developments in information theory, artificial intelligence, and quantum computing, which are generating an enormous number of ethical questions to be answered. Computer and information ethics, then, will have a long and very important role for decades to come.